\documentstyle[preprint,aps]{revtex}

\begin{document}
\title{Elasticity, Stability and Ideal Strength of $\beta $-SiC in plane-wave-based 
{\sl ab initio} calculations}
\author{Weixue Li and Tzuchiang Wang}
\address{LNM, Institute of Mechanics, Chinese Academy of Sciences, Beijing, 100080,\\
China}
\maketitle

\begin{abstract}
On the basis of the pseudopotential plane-wave(PP-PW) method and the
local-density-functional theory(LDFT), this paper studies energetics,
stress-strain relation, stability and ideal strength of $\beta $-SiC under
various loading modes, where uniform uniaxial extension and tension, biaxial
proportional extension are considered along directions [001] and [111]. The
lattice constant, elastic constants and moduli of equilibrium state are
calculated, and the results agree well with the experimental data. As the
four Si-C bonds along directions [111], [$\overline{1}$11], [11$\overline{1}$%
] and [1$\overline{1}$1] are not the same under the loading along [111],
internal relaxation and the corresponding internal displacements must be
considered. We find that, at the beginning of loading, the effect of
internal displacement through shuffle and glide plane diminishes the
difference among the four Si-C bonds length, but will increase the
difference at the subsequent loading, which will result in a crack nucleated
on \{111\} shuffle plane and a subsequently cleavage fracture. Thus the
corresponding theoretical strength is 50.8 GPa, which agrees well with the
recent experiment value, 53.4 GPa. However, with the loading along [001],
internal relaxation is not important for tetragonal symmetry. Elastic
constants during the uniaxial tension along [001] are calculated. Based on
the stability analysis with stiffness coefficients, we find that the
spinodal and Born instabilities are triggered almost at the same strain,
which agrees with the previous molecular dynamics simulation. During biaxial
proportional extension, stress and strength vary proportionally with the
biaxial loading ratio at the same longitudinal strain.
\end{abstract}


\preprint{}

\section{Introduction}

Investigation of stability and ideal strength of materials is always an
attractive issue$^{\text{\cite{frenkel}-\cite{sob}}}$ due to the following
facts: (1). Stability of materials is very important in elasticity theory,
which is related with structural responses in solids, ranging from
ploy-morphism, amorphization, and melting to fracture.$^{\text{\cite{cler95}-%
\cite{mizu94}}}$ (2). The ideal strength of a perfect crystal represents an
upper bound to the actual strength of crystalline materials; (3) The
technology makes it possible to manufacture finer and finer filament, whose
strength will approach the theoretical limit. However, even for the best
whisker materials.$^{\text{\cite{mehan},\cite{petrovic}}}$, the realistic
strength is still far below the predicted theoretical values.

Recent developments of experimental technology see new opportunities of
producing very fine nanorods (NRs) and nanotubes(NTs)$^{\text{\cite{ebbesen}-%
\cite{yang}}}$, and the possibility of measuring their elasticity constants,
strength and toughness,$^{\text{\cite{dai96},\cite{wong}}}$. The typical
measured diameter of NRs is 20-30 nanometer, which can be considered free of
any defects and with the ideal strength. Wong {\sl et al.}$^{\text{\cite
{wong}}}$ measured the Young's modulus and the bending strength of silicon
carbon(SiC) NRs in a recent experiment. It is worth mentioning that
micrometer-scale SiC whisker is widely used to strengthen composite
materials. Thus, it is necessary to have a clear understanding of the
stiffness, stress-strain relation, stability and strength of these nano
sized NRs from the experimental view as well as from the theoretical point
of view.

As is well known, people are used to study the ideal strength of materials
with various models and empirical potentials. Polanyi$^{\text{\cite{polanyi}}%
}$ and Orowan$^{\text{\cite{orowan}}}$ used a model in terms of surface
energy, interplannar space and an appropriate Young's modulus to investigate
the ideal strength. Frenkel$^{\text{\cite{frenkel}}}$ estimated ideal shear
strength $\tau _{max}$ of a solid subjected to deformation of a simple shear
mode. However, oversimplified functions of stress-strain were adopted in
those methods and the functional forms for different materials were somehow
arbitrary. On the basis of stability criteria, Milstein$^{\text{\cite
{milstein71},\cite{milstein73},\cite{mf80}}}$ investigated theoretical
strength of bcc Fe, fcc Ni and Cu with the Morse potential. The ideal
strength is identified with the loss or exchange of stability. This method
helps greatly the investigation of strength and reveals a variety of
interesting and surprising behaviors of materials$^{\text{\cite{mils80a},%
\cite{mils80b},\cite{mils80c},\cite{mils80d}}}$; however, the interatomic
potential they used and obtained by fitting properties of equilibrium state
is inappropriate to be used for the investigation of the stability and
strength of materials, which are essentially of far from being in the
equilibrium state. On the other hand, the density functional theory(DFT)$^{%
\text{\cite{hone64}}}$ , with only the input of atomic position and charge
number , can be used to determine many structural and dynamic properties of
materials under various conditions, including those, which are far from
being in the equilibrium state.

Through a series of comprehensive theoretical and computational studies,
Hill and Milstein$^{\text{\cite{hill75},\cite{hill},\cite{mil82}}}$ have
shown that positive definiteness of internal energy is coordinate dependent
and the stability domain depends on the choices of strain measures; while
Born criteria$^{\text{\cite{born}}}$ are valid only under the zero load.
Based on this idea, Wang {\sl et al.} $^{\text{\cite{wang93},\cite{wang95}}}$
analyzed the onset of instabilities in homogeneous lattice under critical
loading and showed that response of the lattice is no longer a purely
intrinsic property of materials, and depends on the applied load. Starting
with these theories, Li and Wang$^{\text{\cite{li98}}}$ have recently
analyzed the stability and branching of Aluminium under various loading
modes according to the first principle calculations.

Heine and co-workers$^{\text{\cite{heine86},\cite{heine88}}}$ gave a very
extensive set of first principle pseudopotential calculations on the
ploy-types in SiC. However, they gave only the bulk modulus. Lambrecht {\sl %
et al.}$^{\text{\cite{lambrecht}}}$ made a detailed investigation on the
elastic constants, modulus of $\beta $-SiC with the full-potential
linear-muffin-tin-orbital (FP-LMTO) method. In their investigations, the
strength of $\beta $-SiC was obtained approximately by Orowan formula. The
structural properties of $\beta $-SiC(ploy-type 3C) had been also
investigated by the various semi-empirical models, {\sl e.g.} ,
semi-empirical force model$^{\text{\cite{tolpygo}}}$, semi-empirical
interatomic potential$^{\text{\cite{tersoff},\cite{pear84}}}$ and tight
binding approximation$^{\text{\cite{kohy90},\cite{maje87}}}$. Modifying the
Tersoff potential$^{\text{\cite{tersoff}}}$, Tang and Yip$^{\text{\cite{tang}%
}}$ investigated the lattice instability in $\beta $-SiC and simulated the
process of brittle fracture under hydrostatic tension based on the Hill and
Milstein stability theory. The instability mode is the spinodal instability,
and decohesion occurs as spontaneous nucleation of cracking on the $\{111\}$
shuffle planes.

In the present paper, we study the energetics, the elastic constants, the
stress-strain relation, stability and the ideal strength of $\beta $-SiC
with the density functional theory. We consider several loading modes,
uniaxial extension and uniaxial tension along [001] and [111] directions,
and biaxial proportional extension along [001] and [010]. The deformation is
homogeneous and elastic, and the strain can be large. The stress-strain
relations are calculated, and the ideal strength is obtained according to
the stability criteria. Owing to the unequivalence of the four Si-C bonds
under the loading along [111], the internal relaxation must be considered
and the internal displacements be calculated. With the internal
displacements, we discuss the effect of the relaxation and failure modes.
The stability theory of Hill and Milstein$^{\text{\cite{hill75},\cite{hill},%
\cite{mil82}}}$ and Wang $^{\text{\cite{wang95}}}$ are used to discuss
branching and the strength of $\beta $-SiC under the loading along [001].

The present paper is organized as follows. The calculation model is
presented in Sec. II, where we show the formulation of stress, elastic
stiffness coefficients and stability criteria, especially the three loading
modes with the selection of supercell and the numerical precision
illustrated at the end of this Section. The benchmark, including equilibrium
properties and elastic constants, are given in Sec. III. The loading along
[111] direction is presented in Sec. IV, and biaxial proportional extension
is investigated in Sec. V. In Sec. VI, we discuss uniaxial extension and
tension along [001] direction. Summary and conclusion are given in Sec. VII.

\section{Formulation}

Consider an unstressed and unstrained configuration, denoted as ${\bf X}_0$.
It undergoes homogeneous deformation under a uniform applied force, and
changes from ${\bf X}_0$ to ${\bf X=JX}_0$, where ${\bf J}$ is the
deformation gradient or the Jacobian matrix. The associated Lagrangian
strain tensor ${\bf E}$ is: 
\begin{equation}
{\bf E}=\frac 12({\bf J}^T{\bf J-I})  \label{eq:eta}
\end{equation}
where $\ T$ is transpose. The physical strain is: 
\begin{equation}
{\bf e=J-I}  \label{eq:ps}
\end{equation}

For the present deformation, the internal energy {\sl U} is a rotational
invariant and therefore only a function of ${\bf E}$. The second
Piloa-Kirchhoff stress tensor ${\bf T}^{\text{\cite{true60}}}$ is defined
as: 
\begin{equation}
T_{ij}=\frac 1{V_0}\frac{\partial {\sl U}}{\partial E_{ij}}  \label{eq:piloa}
\end{equation}
It is related to Cauchy stress,{\it \ }{\sl i.e}{\rm .} the true stress $%
{\sl \sigma }_{kl}$ by the following equation: 
\begin{equation}
T_{ij}={\bf det}|{\bf J}|{\sl J_{ik}^{-1}J_{jl}^{-1}\sigma _{kl}}
\label{eq:tstress}
\end{equation}
where ${\bf det|J|}$ is the ratio $V/V_0$. With the Cauchy stress, the
applied force can be obtained by multiplying the current transverse area.

At strained state ${\bf X}$, the elastic constants are determined through
the following equation: 
\begin{equation}
C_{ijkl}(X)=\frac 1{V(X)}(\frac{\partial ^2{\sl U}}{\partial
E_{ij}^{^{\prime }}\partial E_{kl}^{^{\prime }}}\mid _{E^{^{\prime }}=0})
\label{eq:elast}
\end{equation}
where ${\bf E}^{\prime }$ is Lagrangian strain around the state ${\bf {\sl X}%
}$. These elastic constants are rotational invariant and symmetric with
interchange of indices $i\leftrightarrow j$, $k\leftrightarrow l$ and $%
(ij)\leftrightarrow (kl)$, which are often expressed in the condensed Voigt
notation.

To analyze the stability, the elastic stiffness coefficient ${\bf B}^{\text{%
\cite{wang95}}}$ is introduced as follows: 
\begin{equation}
B_{ijkl}=C_{ijkl}+\frac 12(\delta _{ik}\sigma _{jl}+\delta _{jk}\sigma
_{il}+\delta _{il}\sigma _{jk}+\delta _{jl}\sigma _{ik}-2\delta _{kl}\sigma
_{ij})  \label{eq:stiff}
\end{equation}
From this definition, we can see that ${\bf B}$ does not posses $%
(ij)\longleftrightarrow (kl)$ symmetry generally. The system may be unstable
when 
\begin{equation}
{\bf det|B|}=0  \label{eq:stability2}
\end{equation}
for the first time.

The following loading modes are considered:

(i) Uniaxial Extension. 
\begin{equation}
e_{ij}=e\delta _{i3}\delta _{j3}\text{,}\ \ \ \ \ \ \ i,j=1,2,3
\label{eq:ud}
\end{equation}

(ii) Uniaxial Tension. 
\begin{equation}
\sigma _{ij}=\sigma \delta _{i3}\delta _{j3}\text{, \quad }\ i,j=1,2,3
\end{equation}
For a given longitudinal strain, let the transverse lattice contract or
dilate to make the total energy approach minimum, which corresponds zero
stress(traction) on lateral faces. For crystal symmetry, the transverse
contraction is the same at two perpendicular transverse directions, so 
\begin{equation}
e_{11}=e_{22}=-\lambda e_{33}  \label{eq:tc}
\end{equation}

(iii) Biaxial Proportional Extension 
\begin{equation}
e_{22}=\alpha e_{33}\not{=}0\ \ \ \ e_{ij}=0\ \ \ others  \label{td}
\end{equation}

The total energy calculations are carried out with {\sl ab initio}
pseudopotential plane-wave program package Fhi96md.$^{\text{\cite{fhi96md}}}$
By means of the mechanism of Hamman $^{\text{\cite{hamman}}}$ and Troullier$%
^{\text{\cite{troullier}}}$, the soft first principle pseudopotential$^{%
\text{\cite{fuchs},\cite{gonze}}}$ is generated, where the local density
approximation(LDA) with the exchange and correlation energy functional
developed by Perdew and Zunger$^{\text{\cite{perdew}}}$ is used. Two
supercells are designed in our calculations: one is the 8-atom supercell for
the equilibrium properties, the loading along [001] and biaxial extension
along [010] and [001]. The other one is the 6-atom supercell for the loading
along [111]; in this case, the stacking consequence is Si-C-Si-C-Si-C. There
exist two types of \{111\} plane, between Si and C atoms, corresponding to
the well known glide and shuffle planes. The glide plane cuts three Si-C
bonds out of four, and the shuffle plane cuts the remaining Si-C bond. For
numerical differential feature of stress and elastic constants, the
precision must be considered carefully. The size of carbon atom is so small
that a high cut-off energy is required. Our test shows that $E_{cut}=80Ry$
has also given excellent results.. The k-space mesh is $6\times 6\times 6$
for the 8-atom supercell and $8\times 8\times 4$ for the 6-atom supercell in
order to keep same precision.

\section{Equilibrium Properties}

As the benchmark, we have calculated the lattice constant, elastic constants
and moduli of $\beta $-SiC of equilibrium. For symmetry of $\beta $-SiC
(zincblende structure), there exist three independent elastic constants, 
{\sl i.e.} $C_{11}$, $C_{12}$, $C_{44}$. The total energy of $\beta $-SiC is
calculated under the applied hydrostatic, uniaxial deformation, and trigonal
strain. Owing to the lattice feature of zincblende structure, which includes
two fcc lattices with a relative displacement along [111], the symmetry of
center inversion is lost. Four Si-C bonds along directions [111], [1$%
\overline{1}$1], [11$\overline{1}$] and [$\overline{1}$11] are not
equivalent under the case of trigonal strain. The internal atomic position
must be fully relaxed, and the internal displacement$^{\text{\cite{cousins}}%
} $, which refers to the relative displacement of two sublattices beside the
displacement from the macroscopic strain, will take place. Our results are
presented in Table I. From this table, we find, our results agree well with
the experimental data and the previous first principle and semi-empirical
calculations. The value of $C_{44}$ without relaxation, 270 GPa, is already
in good agreement with the experimental data and better than other
theoretical calculations. The relaxed value, 254 GPa, is almost the same as
the experiment value. Value of the anisotropy ${\bf A}$ is also satisfactory.

Based on the representation surface$^{\text{\cite{nye}}}$, the moduli along
a certain direction can be obtained. The Young' modulus along directions
[111] and [001] are 554 GPa and 338 GPa, respectively. Lambrecht {\sl et al.}
$^{\text{\cite{lambrecht}}}$ obtained 603 GPa and 362 GPa. Petrovic {\sl et
al.} $^{\text{\cite{petrovic}}}$ measured Young's modulus of $\beta $-SiC
whisker, with an averaged value of 578 GPa with $\pm 10\%$ scattering.
Applying the equation of the cantilever beam, Wong {\sl et. al.}$^{\text{%
\cite{wong}}}$ measured Young's modulus of [111]-oriented SiC nanorod, which
are 610GPa and 660 GPa, corresponding to the 23.0-nm-diameter and
21.5-nm-diameter SiC NRs. The agreement is good.

With orientation averages, the moduli of isotropic materials can be
obtained. Two average methods , namely Reuss averages$^{\text{\cite{reuss}}}$
($E_R$ and $G_R$ ) and Voigt averages$^{\text{\cite{voigt}}}$, (${\sl E_V}$
and ${\sl G_V}$ ) are adopted. According to the theory of Hill$^{\text{\cite
{hill52}}}$, the physical averages, here denoted by subscript ${\sl a}$, are
the intermediate between the Reuss and Voigt averages. With these
considerations, Young' modulus and shear modulus of isotropic $\beta $-SiC
are given as follows: 
\[
\begin{array}{lcc}
& E_a & \approx 448GPa\pm 2.2\% \\ 
& G_a & \approx 192GPa\pm 2.6\%
\end{array}
\]
Compared with the previous first principle$^{\text{\cite{lambrecht}}}$ and
semi-empirical $^{\text{\cite{tolpygo},\cite{tersoff}}}$ calculations, our
results agree better with the experimental values. These results confirm the
conclusion of Lambrecht: the random orientation hypothesis applies well to
the ceramic samples. The average Possion ratio $\nu _a$, 0.17, is close to
the experimental value. The small Possion ratio of $\beta $ Si-C, as
compared with other materials, {\sl e.g.} Aluminum 0.347, demonstrates its
high stiffness .

\section{Loading Along Direction [111]}

In this loading direction, two loading modes are investigated: uniaxial
extension and uniaxial tension. At the latter case, the transverse lattices
contract in order to approach the energy minimum. With regard to the loss of
symmetry of center inversion, the four Si-C bonds along directions [111], [1$%
\overline{1}$1], [11$\overline{1}$] and [$\overline{1}$11], are not
equivalent under [111] loading. The internal relaxation and lateral
contraction must be considered. In our calculations, by using the lattice
constant at room temperature, the internal relaxation is carried out after
the transverse strains are obtained. The length of Si-C bond is 3.5673
(Bohr) under zero loading. As the loading is along [111] direction, the
variation of bond length along the [111] direction will be more significant
than that of the other three bonds. Fig.1 shows the energy, stress and force
under uniaxial extension and uniaxial tension with or without internal
displacement.(Without other statement, the strain, force and stress, given
in figures, are the physical strain, applied force and Cauchy stress.) The
corresponding transverse strain and internal displacement are given in Fig.
2 and Fig 3.

At the beginning of loading, as compared with the corresponding Si-C bond
length without the internal relaxation, the difference among the four Si-C
bond lengths is small, and no marked effect of relaxation and internal
displacement through both the shuffle and glide plane is shown. This
phenomenon, shown in Fig.1, is obvious for $\beta $-SiC, a kind of high
stiffness and low Possion ratio covalent material. The strain energy curves
of the three loading modes, {\sl i.e. }uniaxial extension and uniaxial
tension with or without internal relaxation are almost the same. Despite the
fact that the stress and force of uniaxial tension is smaller than the
uniaxial extension for the relaxation, they are still similar in these
loading modes. Based on Kleinman's$^{\text{\cite{kleinman}}}$ discussion on
Silicon with [111] strain, the internal strain tends to keep the bond length
along the four unchanged unequivalent [111] directions. In our calculations,
the internal displacement of atom along [111] direction, which is through
the shuffle plane, is negative, and that of the remaining three Si-C bonds
along the [1$\overline{1}$1], [11$\overline{1}$] and [$\overline{1}$11]
directions, through the glide plane, is positive. This means that the effect
of relaxation always tends to diminish the difference of four Si-C bond
lengths. With the increase of the longitudinal strain, the internal
displacement through the shuffle plane becomes positive from negative, and
that through the glide plane moves from positive to negative while
approaching zero at the same point ( $e_z=0.105$). The internal
displacements for the two cases have the same magnitude but with the
opposite sign. The details can be found in Fig.3. During the whole uniaxial
tension, the magnitude of transverse strain increases monotonically.

With further increase of the longitudinal strain, the strain energy of
uniaxial extension and of uniaxial tension without internal displacement
still have approximately the same value. However, both the internal
displacements through the shuffle plane and the glide plane change their
signs ( the symmetry is still hold). The uniaxial tension curve with
internal relaxation softens quickly, and the shape of tensile curve changes
dramatically. An energy plateau presents, and material becomes unstable. On
the basis of the stress curve, the maximum stress of uniaxial tension with
internal relaxation, namely the theoretical strength $\sigma ^{th}$, is
obtained and equal to 50.8 GPa. The corresponding critical macroscopic
strain and internal displacement is 0.144 and 0.082 (Bohr), the Si-C bond
length along [111] is 4.163 (Bohr). With the modified Tersoff potential,
Tang and Yip$^{\text{\cite{tang}}}$ analyzed the brittle fracture of $\beta $%
-SiC under hydrostatic tension by molecular dynamics. They found that the
mode of instability of $\beta $-SiC was the spinodal instability, and the
corresponding critical strain and pressure were 0.153 and 37.0GPa.
Therefore, both the first principle and empirical potential calculations
gave a similar critical bond-length of $\beta $-SiC along [111].

Our result agrees well with the experimental value, given by Wong {\sl et al.%
} $^{\text{\cite{wong}}}$ , 53.4 GPa, obtained for [111]-oriented SiC
nanorod. This agreement also means that no other branching and instability
modes exist during the uniaxial loading before it reaches the inflexion of
energy-strain curve. It is worth while pointing out that the experimental
strength measured is the bending strength. The tensile and bending strengths
are comparable to $\beta $-SiC whisker$^{\text{\cite{macmillan},\cite
{petrovic}}}$ and are also expected to be similar to $\beta $-SiC nanorod.$^{%
\text{\cite{wong}}}$ Petrovic {\sl et al.}$^{\text{\cite{petrovic}}}$
measured the tensile strength of $\beta $-SiC whisker and their result is
23.74 GPa, which is far smaller than our theoretical calculation and Wong's
experiment values for defects. Lambrecht {\sl et al.} $^{\text{\cite
{lambrecht}}}$ calculated the tensile strength by Orowan expression with
[111] surface energy, and the result is 30 GPa. With the similar formula, Op
Het Veld and Veldkamp$^{\text{\cite{op}}}$ obtained the theoretical cleavage
strength 46.3 GPa, which is close to ours. The detailed comparison can be
found in Table 2.

After $e_z>0.105$, the internal displacement through the shuffle plane
becomes positive, and that through the glide plane becomes negative. The
distance between the atom through the shuffle plane along [111] increases,
and that through the glide plane decreases, and a crack nucleates on the
\{111\} shuffle plane. The internal displacements through the shuffle and
glide plane at the critical strain are, respectively, 0.082 and -0.082. With
further increase of the longitudinal strain, the internal displacements of
the shuffle and glide planes also increase quickly. This positive and
negative increase of internal displacements of shuffle and glide planes will
result in a dramatic cleavage on the \{111\} shuffle plane and the mixing of
Si and C atoms through the glide plane. The cleavage on \{111\} shuffle
plane can be partly attributed to the lower surface energy than that of the
\{111\} glide plane.$^{\text{\cite{hill75},\cite{tang},\cite{oshc}}}$ These
results agree well with the previous molecular dynamics simulation.$^{\text{%
\cite{tang}}}$

\section{Biaxial Proportional Extension}

To consider only the biaxial proportional extension, this section deals with
the extension along directions [010] and [001], and not the internal atomic
and volume relaxation. The strain ratio between [010] and [001] is 0.25,
0.5, 0.75 and 1. The results are shown in Fig.4. With the increase of the
ratio, the energy, stress and maximum stress will increase at the same
longitudinal strain accordingly. However, the critical strain is similar for
different proportional loading modes.

\section{Loading along Direction [001]}

In this section, we consider the uniaxial extension and uniaxial tension
along direction [001]. The reference state is the state with the theoretical
lattice constant. Symmetry of crystal under this loading mode is tetragonal.
Unlike the loading along [111], the deformation of the four Si-C bonds in
this loading mode is the same and the four bonds are equivalent. There will
not be any internal displacements and the internal relaxation can be
neglected during the loading. We have made calculations at several strain
with or without internal relaxation and found that the value of transverse
strain at a specific longitudinal strain is the same. Our results are given
in Fig.2, Fig.5

From Fig.5a, we can see that both of the strain energy for uniaxial
extension and tension increase with the increase of the longitudinal strain.
The strain energy of uniaxial extension is always larger than that of the
uniaxial tension, as is expected. However, the energy difference between two
loading modes is small, same as with the [111] loading. At a larger strain,
the energy difference becomes even smaller. The applied force and stress of
uniaxial tension are lower than those of the uniaxial extension for the
triaxial stress state at the beginning and higher than them at the
subsequent loading ( this phenomenon will be explained later). Just like
Fig.5a, the difference of applied force and stress between these two loading
modes is not significant.

In order to obtain the ideal strength and analyze the stability under
uniaxial tension, we calculate the elastic constants and derive the
stability criteria based on the stiffness coefficients. With the tetragonal
symmetry, the number of independent elastic constants is reduced to six: $%
C_{33}$, $C_{12}$, $C_{13}=C_{23}$, $C_{11}=C_{22}$, $C_{44}=C_{55}$ and $%
C_{66}$; all the other $C_{ij}$ are equal to zero. With equation \ref
{eq:stiff}, \ref{eq:stability2} and \ref{eq:ul}, we write the stability
criteria as follows: 
\begin{eqnarray}
(C_{33}+\sigma )(C_{11}+C_{12})-2C_{13}(C_{13}-\sigma ) &\geq &0
\label{eq:uls} \\
C_{11}-C_{12} &\geq &0 \\
C_{44}+\frac 12\sigma &\geq &0 \\
C_{66} &\geq &0
\end{eqnarray}
The first one involves the vanishing of bulk modulus, and is referred as
spinodal instability. The second instability involves symmetry breaking
(bifurcation) with the volume conservation; it may be identified as the
tetragonal shear breaking and referred as Born instability. The third and
fourth are two distinct shear deformation instabilities. Six strains are
designed to calculate the independent elastic constants and given as follows:

\begin{eqnarray*}
e_{11} &=&e_{22}=\delta \text{, }e_{ij}=0\text{, i, j=1,2,3} \\
e_{11} &=&-e_{22}=\delta \text{, }e_{ij}=0\text{, i, j=1,2,3} \\
e_{11} &=&e_{33}=\delta \text{, }e_{ij}=0\text{, i, j=1,2,3} \\
e_{11} &=&-e_{33}=\delta \text{, }e_{ij}=0\text{, i, j=1,2,3} \\
e_{12} &=&e_{21}=\delta \text{, }e_{ij}=0\text{, i, j=1,2,3} \\
e_{23} &=&e_{32}=\delta \text{, }e_{ij}=0\text{, i, j=1,2,3}
\end{eqnarray*}
{In each case, the domain of strain is [0,0.02]. The results are shown in
Fig.6}

During uniaxial tension, the variation of mechanical properties on
transverse section is comparatively small, and the corresponding elastic
constants, {\sl i.e.}, $C_{11}$, $C_{12}$, $C_{66}$, keep positive, as shown
in Fig.6a. However, the elastic constants related to longitudinal strain
change dramatically and even become negative at large strain, {\sl e.g.} $%
C_{13}\leq 0$ when $e_{33}\geq 0.184$, $C_{33}\leq 0$ when $e_{33}\geq 0.352$%
. Because $C_{13}\leq 0$ leads to negative Possion ratio, the transverse
section will expand with the increase of the longitudinal strain. This
phenomenon is also shown in Fig.2, the transverse strain varies $e_{11}$
proportionally with $e_{33}$ when $e_{33}\geq 0.20$. Negative Possion ratio
had also been investigated by Milstein {\sl et al} $^{\text{\cite{milstein71}%
,\cite{milstein73},\cite{mils80b},\cite{mils79}}}$. However, in their
papers, negative Possion ratio only occurs at branching or unstable points,
and the materials investigated are monatomic metal materials Fe and Ni. In
the present calculations, it is surprising that $\beta $-SiC, a non-metal
and two component crystal, is still stable at this negative Possion ratio,
shown in Fig.6b. The same result is obtained when calculations with internal
atomic relaxation are implemented. This phenomenon must be related with the
bond nature of $\beta $-SiC. The charge transfer$^{\cite{harr80}}$ and ionic
component$^{\cite{verm66}}$ of $\beta $-SiC will affect the mechanical
response. The detailed analysis of electric structure should be carried out
and further investigation is necessary. The lattice transverse expansion
leads to much quicker increase of the force and stress of uniaxial tension
than that of uniaxial extension and the values of the previous force and
stress will be higher than those of uniaxial extension at a large strain.

On the basis of the stability criteria, we have found that the spinodal and
Born instabilities are triggered almost at the same strain 0.37 with the
transverse strain -0.0137. The corresponding strength, 101.3 GPa, which is
almost twice of that for [111]- oriented SiC nanorod 53.4 GPa$^{\text{\cite
{wong}}}$ for its smaller interplannar distance, is obtained. At this
critical strain, the elongation strain along [111] direction is 0.129,
comparable with the critical strain 0.144 under uniaxial lading along [111].
Tang and Yip$^{\text{\cite{tang}}}$ investigated the instability of $\beta $%
-SiC under hydrostatic tension with stiffness coefficient, and found that
failure mode of $\beta $-SiC is spinodal instability. This was proved by
their molecular dynamics simulations, and the nucleation of cracking on the $%
\{111\}$ plane and decohesion were revealed. They also showed that a shear
instability is triggered by the spinodal instability. All of these results
are same with our first principle calculations.

\section{Summary and Conclusion}

On the basis of the DFT total energy calculation and stability theory, we
give a detailed investigation of mechanical properties of two-atomic
constituent materials $\beta $-SiC: energetics, elasticity, stress-strain
relations, stability and strength under different loading modes and
directions. The results are satisfactory.

Owing to the unequivalence of the four Si-C bonds under the uniaxial tension
along [111], the relaxation must be implemented and internal displacements
be calculated. The internal displacements along the [111] direction and
other three directions namely [$\overline{1}$11], [1$\overline{1}1$] and [1$%
\overline{1}$1] have the same magnitude but the opposite sign. At the
beginning of loading, the effect of relaxation is not significant and tends
to diminish the difference of the four Si-C bond lengths. However, it
becomes important at the subsequent loading and results in a crack nucleated
on \{111\} shuffle plane, while the Si atom and C atom through the glide
plane approach each other. The failure in this loading modes is of cleavage
fracture. These conclusions are consistent with the previous molecular
dynamics simulations. The theoretical strength obtained agrees well with the
experimental data.

Under loading along [001], the four Si-C bonds are equivalent for the
tetragonal symmetry and the relaxation can be neglected. The strain energy,
applied force and stress are similar despite two distinct loading modes,
namely, uniaxial tension and extension. During the uniaxial loading along
[001], the spinodal and Born instabilities are triggered almost at the same
strain. Previous molecular dynamics investigation revealed similar facts.
Owing to the smaller interplannar distance, the corresponding ideal
strength, 101 GPa, which is much higher than the theoretical strength for
the loading along [111] and the experimental data, is obtained; however, the
Si-C bond length for loading along [001] and [111] at the critical strain is
close. There exists some stable range when the Poisson ratio is negative,
and these phenomena are related to the unique bond nature of $\beta $-SiC, a
kind of non-typical covalent material, which permits charge transfer. A
detailed analysis of electronic structure will brought up for further
investigation.

\acknowledgements

This work was supported by the National Natural Science Foundation of China
(Grant No.19704100) and National Natural Science Foundation of the Chinese
Academy of Sciences (Grant No. KJ951-1-201). One of authors W. X. Li wishes
to thank Prof. D. S. Wang for his helpful discussion and encouragement.
Parts of the computations are done at the super-parallel computer of the
Network Information Center of the Chinese Academy of Sciences. We are also
grateful to Prof. K. Wang for checking the manuscript.

\newpage 
\begin{table}[tbp]
\caption{ Equilibrium and elastic modulus of $\beta$-SiC. PP-PW, present
pseudopotential plane wave calculations; FP-LMTO, Lambrecht (Ref.26); CKH,
Churcher, Kunc and Heine (Ref.24); Tolpygo (Ref. 22); Tersoff (Ref.23);
Exp., expermental values as indicated by footnotes. The length unit is bohr
radius, and modulus is GPa, the anisotropy A=2$C_{44}/(C_{11}-C_{12})$. }
\begin{center}
\begin{tabular}{lcccccc}
& PP-PW & FP-LMTO & CKH & Tolpygo & Tersoff & Exp. \\ \hline
$a_0$ & 8.166 & 8.154 & 8.145 &  & 8.164 & 8.238$^a$ \\ 
${\sl B}_0$ & 225 & 223 & 224 & 211 & 220 & 225$^b$ \\ 
$C_{11}$ & 405 & 420 &  & 352.3 & 420 & 390$^c$ \\ 
$C_{12}$ & 135 & 126 &  & 140 & 120 & 142$^c$ \\ 
$C_{44}$ & 254(270) & 287 &  & 232 & 260 & 256$^c$ \\ 
A & 1.88 & 1.95 &  & 2.20 & 1.73 & 2.00$^c$ \\ 
$E_{111}$ & 558 & 603 &  & 511 & 560 & 581($\pm$10\%)$^d$ 610$^e$ \\ 
$E_{100}$ & 338 & 362 &  & 272 & 367 &  \\ 
$E_{R}$ & 441 & 476 &  & 378 & 462 &  \\ 
$E_{V}$ & 474 & 516 &  & 424 & 488 &  \\ 
$E_{a}$ & 458 & 496 &  & 401 & 475 & 448$^b$ \\ 
$G_{R}$ & 188 & 208 &  & 157 & 201 &  \\ 
$G_{V}$ & 206 & 231 &  & 182 & 216 &  \\ 
$G_{a}$ & 197 & 219 &  & 169 & 208.5 & 192$^b$ \\ 
$\nu_a$ & 0.173 & 0.146 &  & 0.201 & 0.150 & 0.168$^b$%
\end{tabular}
\end{center}
\par
$^a$ Landolt and B\"{o}rnstein(\cite{landolt}) \hspace{0.2cm} $^b$ Carnahan(%
\cite{carnahan}) \hspace{0.2cm} $^c$ Obtained from sound velocities (\cite
{feldman}) \newline
$^d$ Experimental values from whisker(\cite{petrovic}) \hspace{0.3cm} $^e$
Experemental values from nanorods(\cite{wong})
\end{table}

\begin{table}[tbp]
\caption{The Young's modulus and strength compared to other theoretical
calculations and experiments; Here, ts, bs and cs means the tensile
strength, the bending strength and the cleavage strength. The unit is GPa.}
\begin{center}
\begin{tabular}{c|ccccccccc}
& ${\sl E}$ & $\sigma_{ts}$ & ${\sl E}^{a}$ & $\sigma_{ts}^{a}$ & $%
\sigma_{cs}^{b}$ & ${\sl E}^{c}$ & $\sigma_{bs}^{c}$ & ${\sl E}^{d}$ & $%
\sigma_{ts}^{d}$ \\ \hline
$[111]$ & 558 & $50.4(0.144)$ & 603 & 30 & 46.3 & 610 & 53.4 & 580$\pm10\%$
& 23.74 \\ 
$[001]$ & 338 & $101(0.37)$ & 362 &  &  &  &  &  & 
\end{tabular}
\end{center}
\par
$^a$ From FP-LMTO and Orwan expression cite{lambrecht} \hspace{0.3cm} $^b$
From Orwan expression $\cite{op}$ \newline
$^c$ Experimental values from nanorods \cite{wong} \hspace{0.3cm} $^d$
Experimental values from whiskers \cite{petrovic}
\end{table}

\newpage 
\begin{figure}[tbp]
\caption{ (a) The calculated strain energy under the uniaxial extension and
uniaxial tension with or without internal relaxation; (b) Applied force and
stress under uniaxial tension with or without relaxation.}
\end{figure}

\begin{figure}[tbp]
\caption{The transverse strain of uniaxial tension along directions [001]
and [111].}
\end{figure}

\begin{figure}[tbp]
\caption{The internal displacement through shuffle and glide plane under
[111] uniaxial tension. The club is for the shuffle plane, and the square is
for the glide plane.}
\end{figure}

\begin{figure}[tbp]
\caption{ The calculated strain energy(a) and applied stress(b) during
biaxial proportional extension with difference ratio along [010]($e_{22}$)
and [001]($e_{33}$) direction.}
\end{figure}

\begin{figure}[tbp]
\caption{The calculated strain energy (a), applied force and stress (b) of
uniaxial extension and uniaxial tension along direction [001].}
\end{figure}

\begin{figure}[tbp]
\caption{ The calculated elastic constants (a) and stability (b) during
uniaxial tension along direction [001].}
\end{figure}

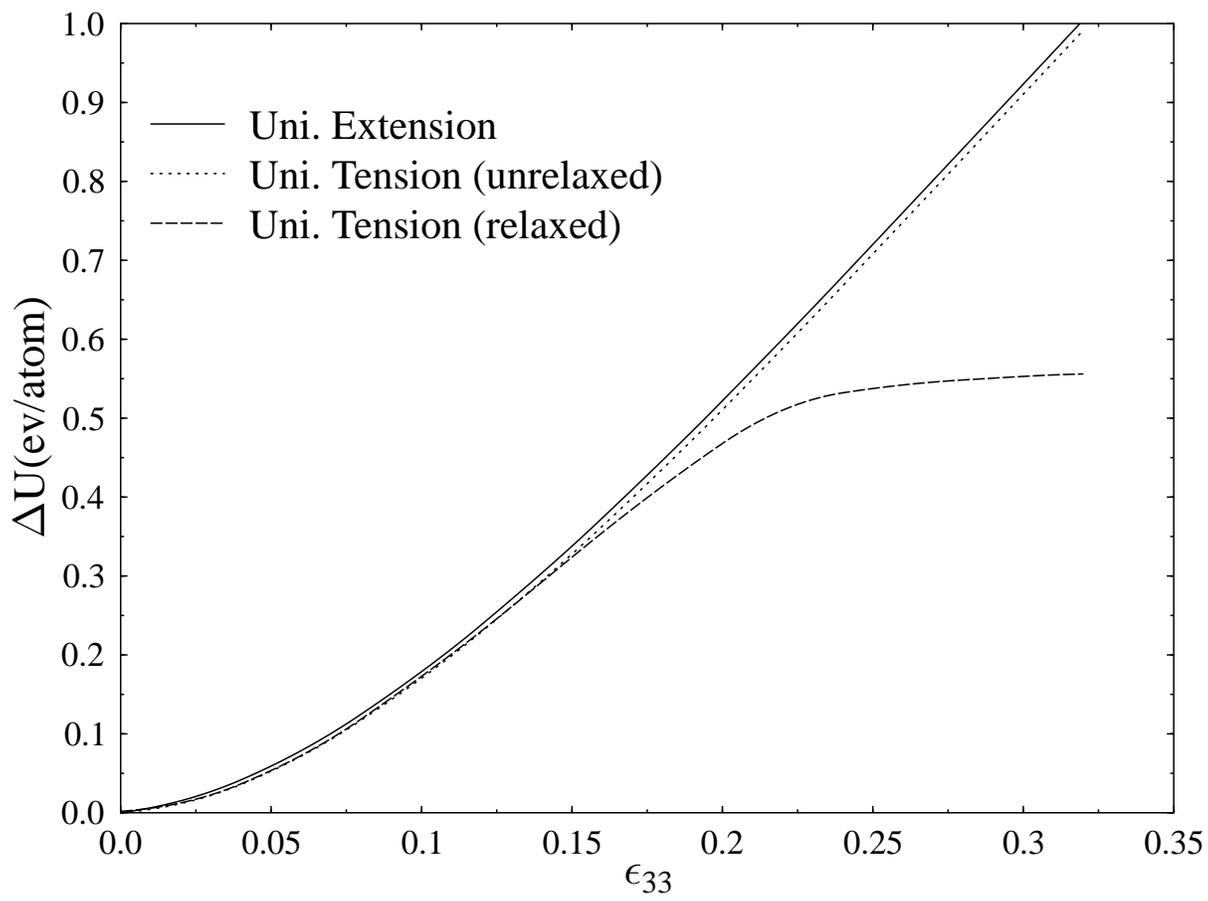
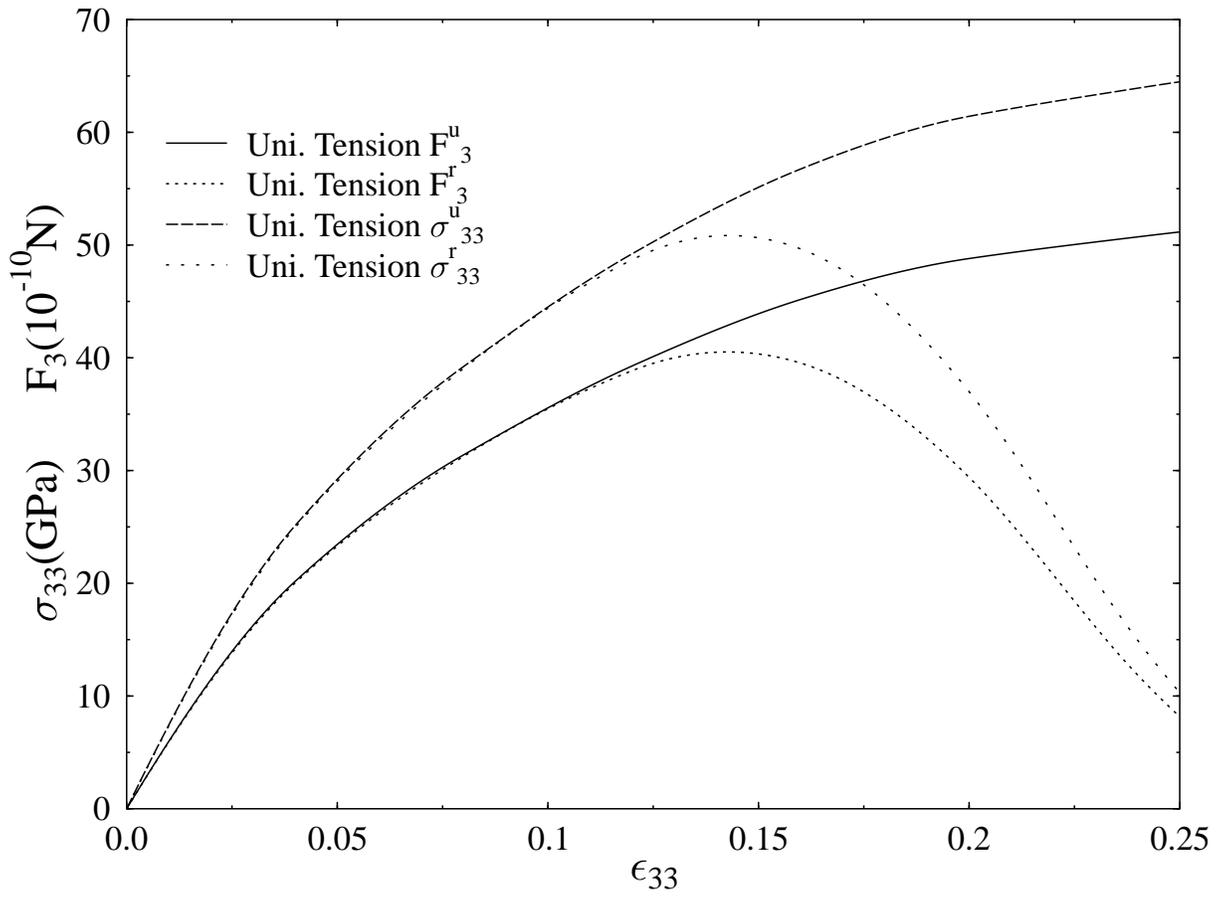
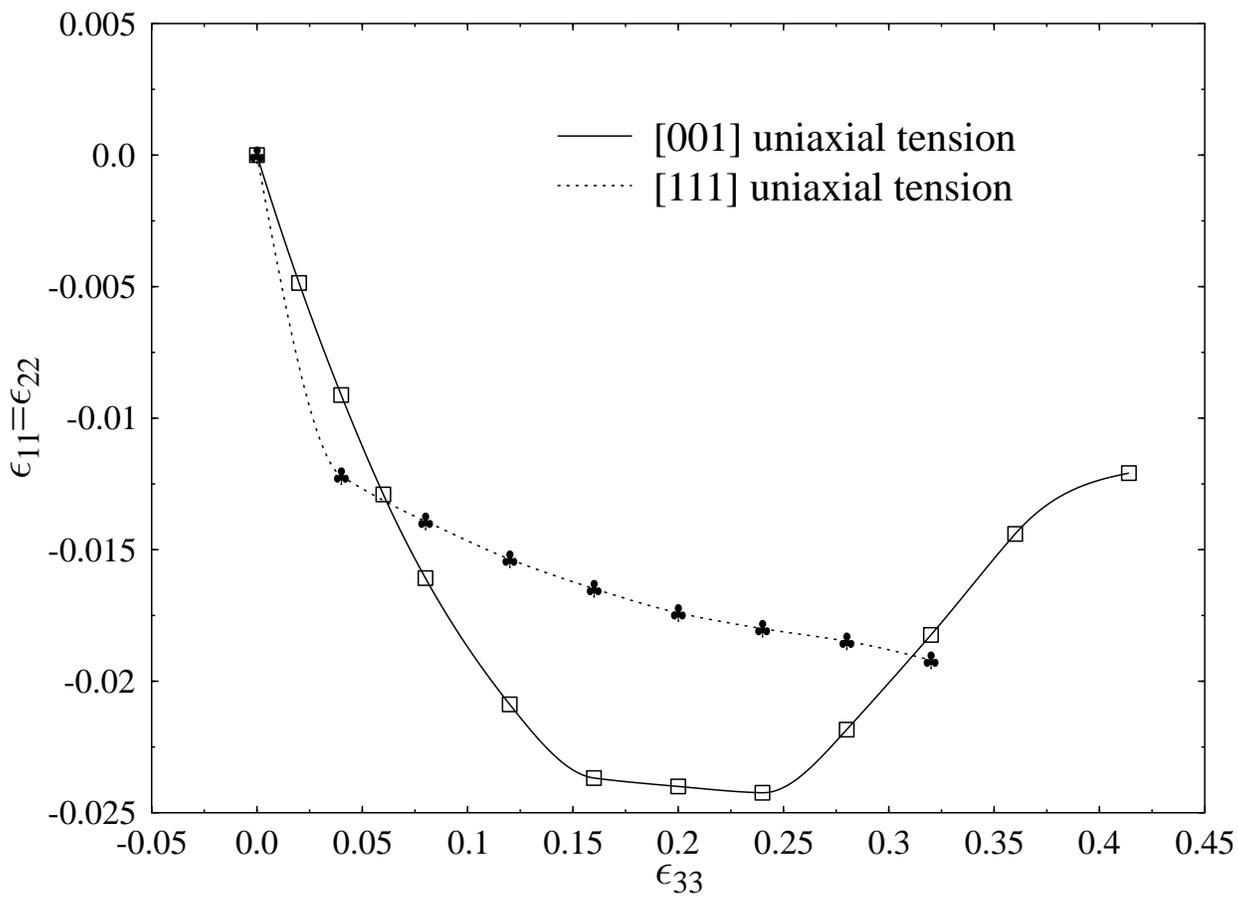
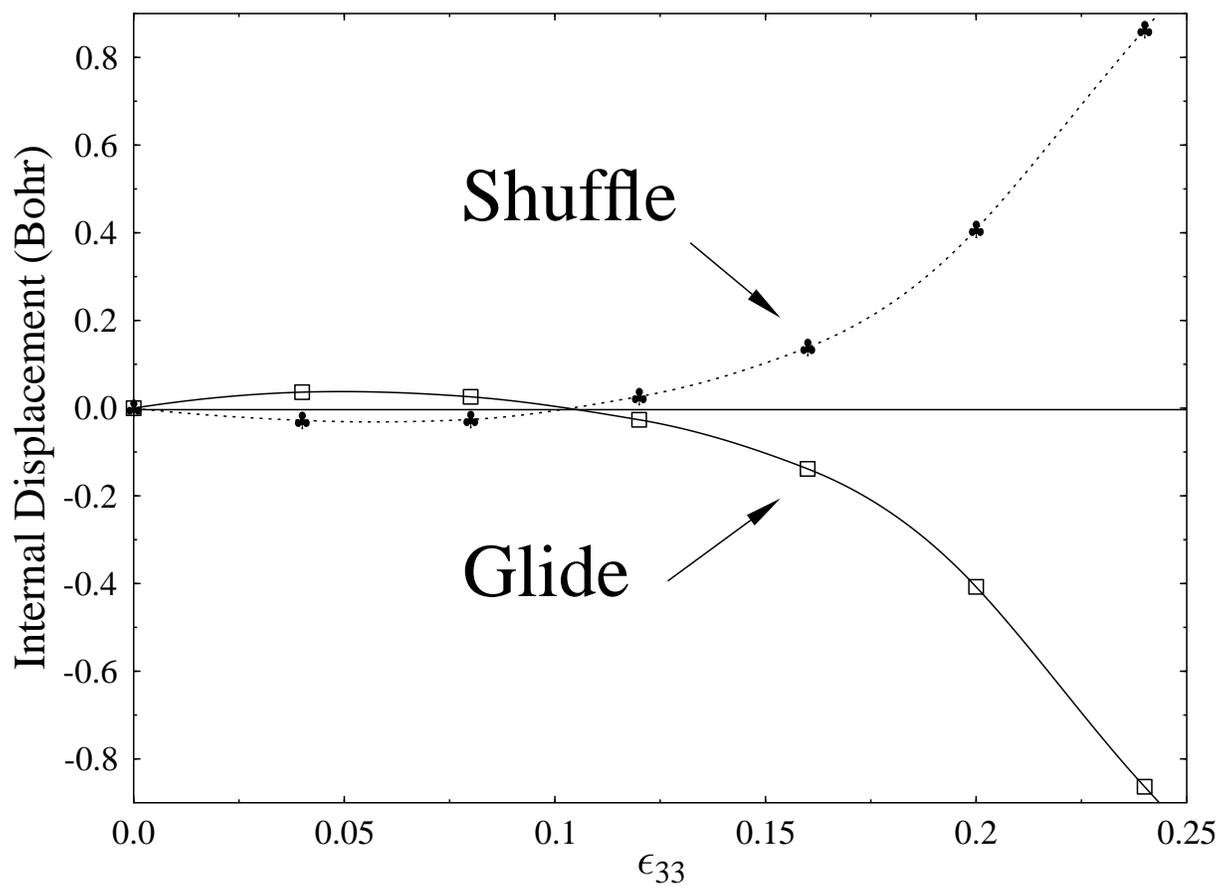
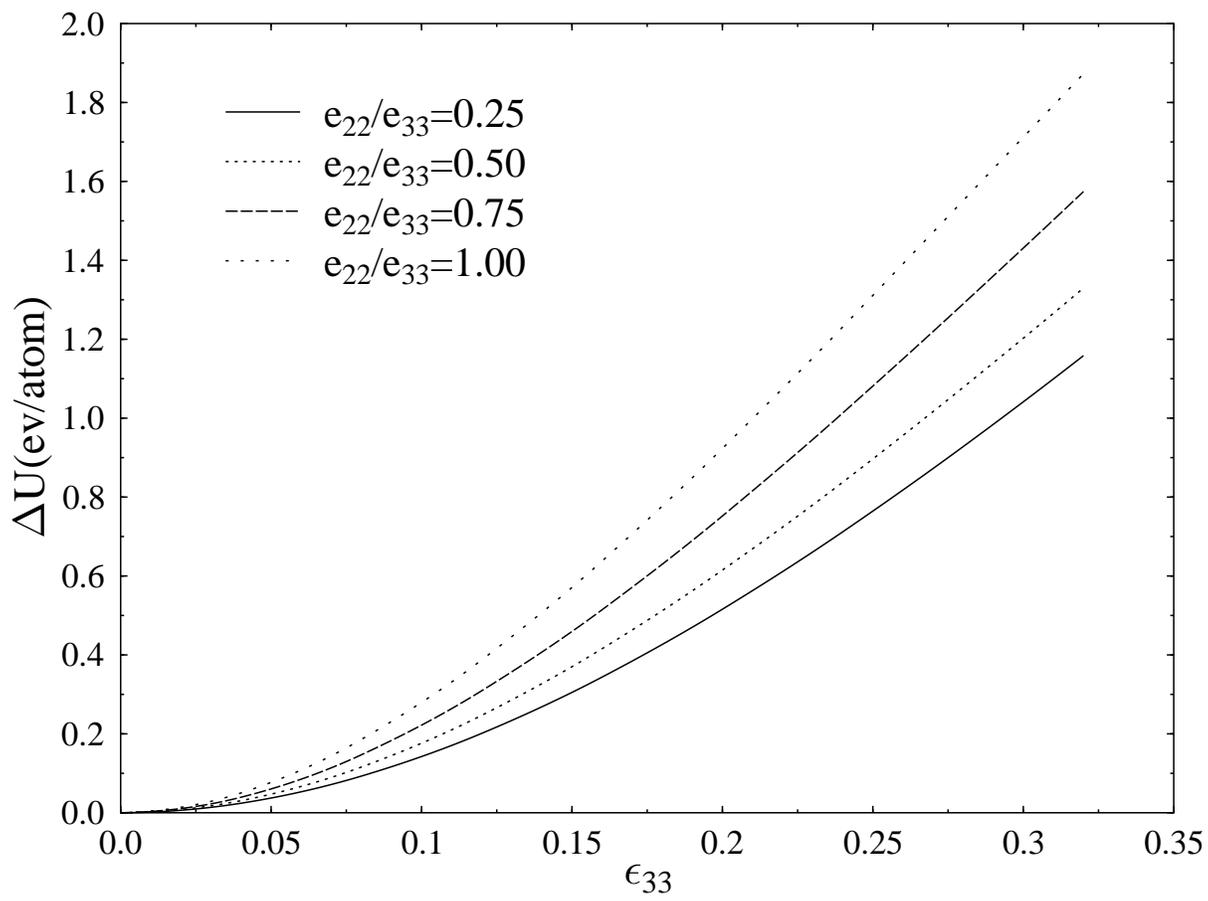
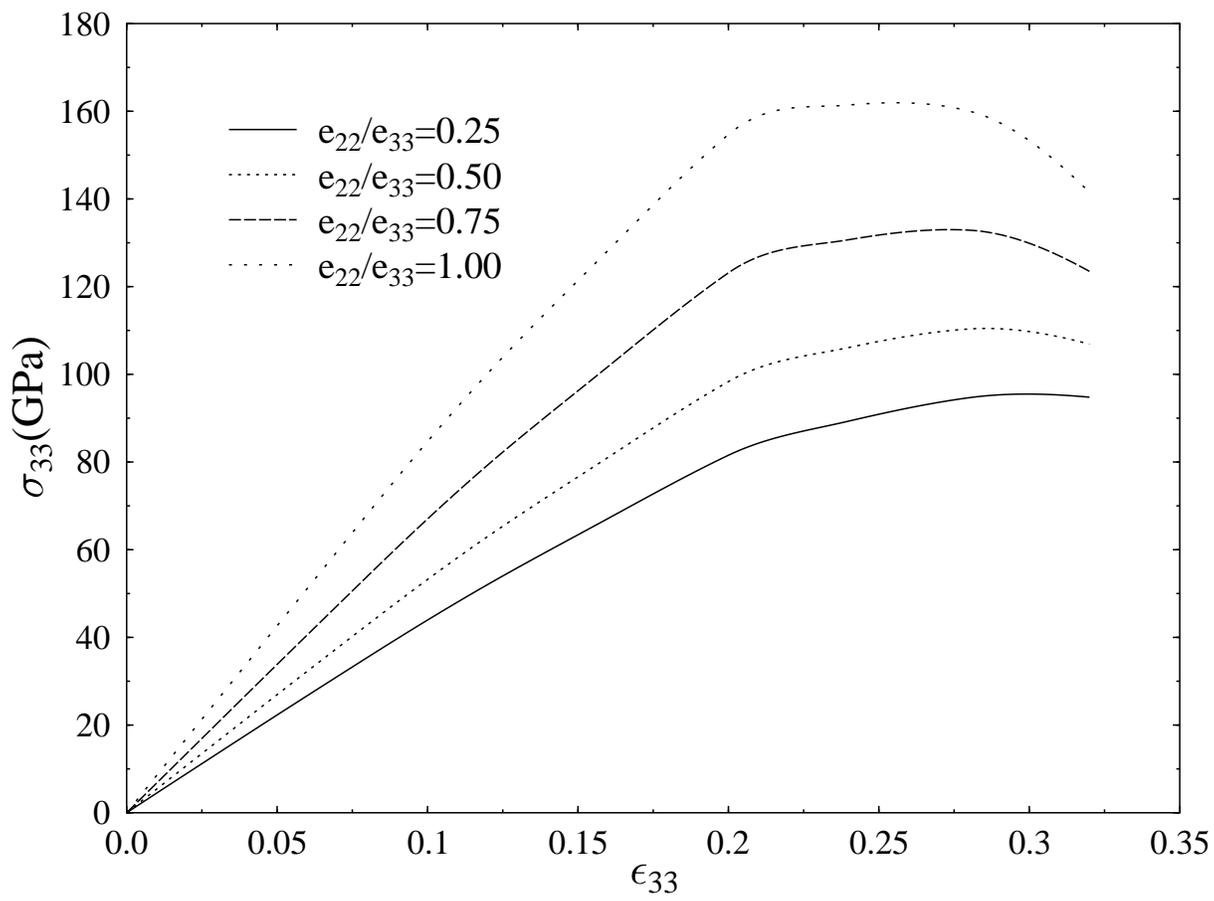
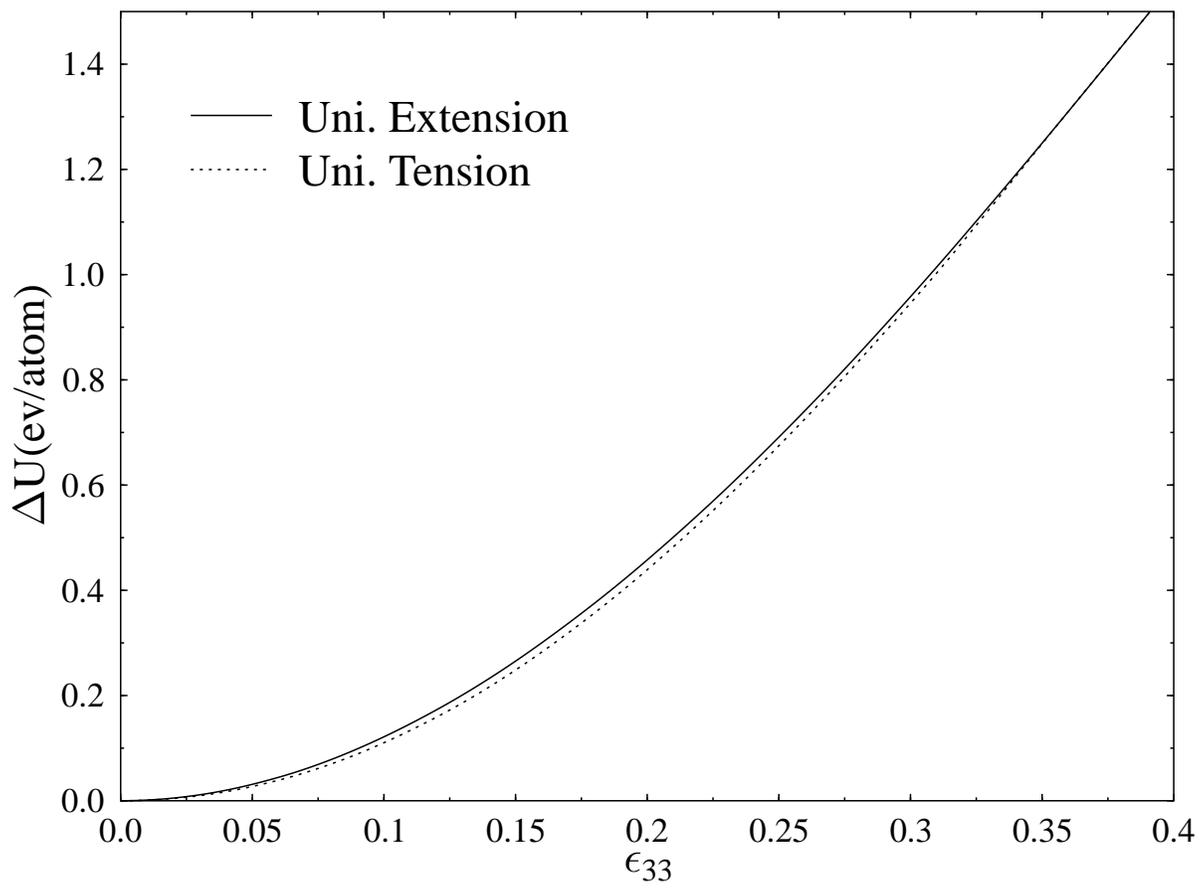
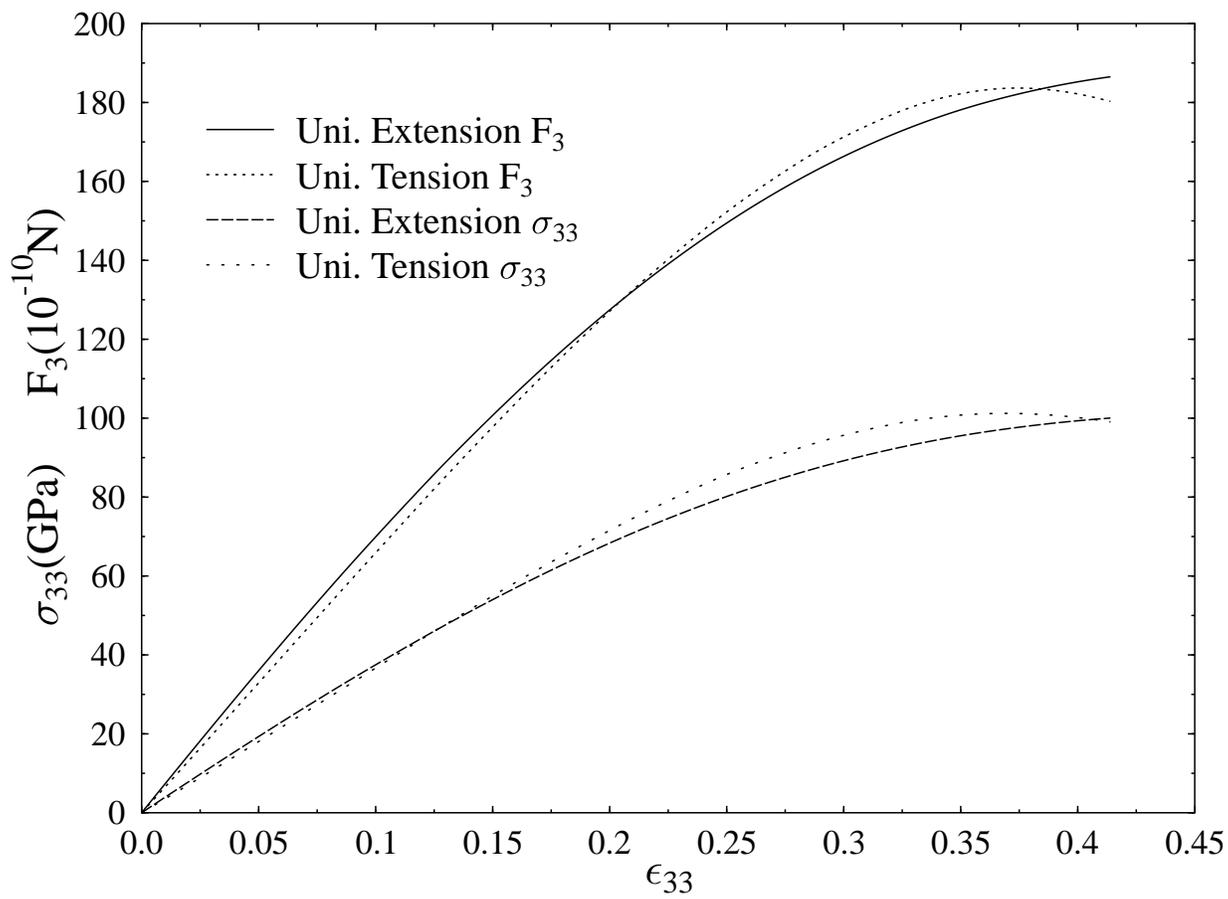
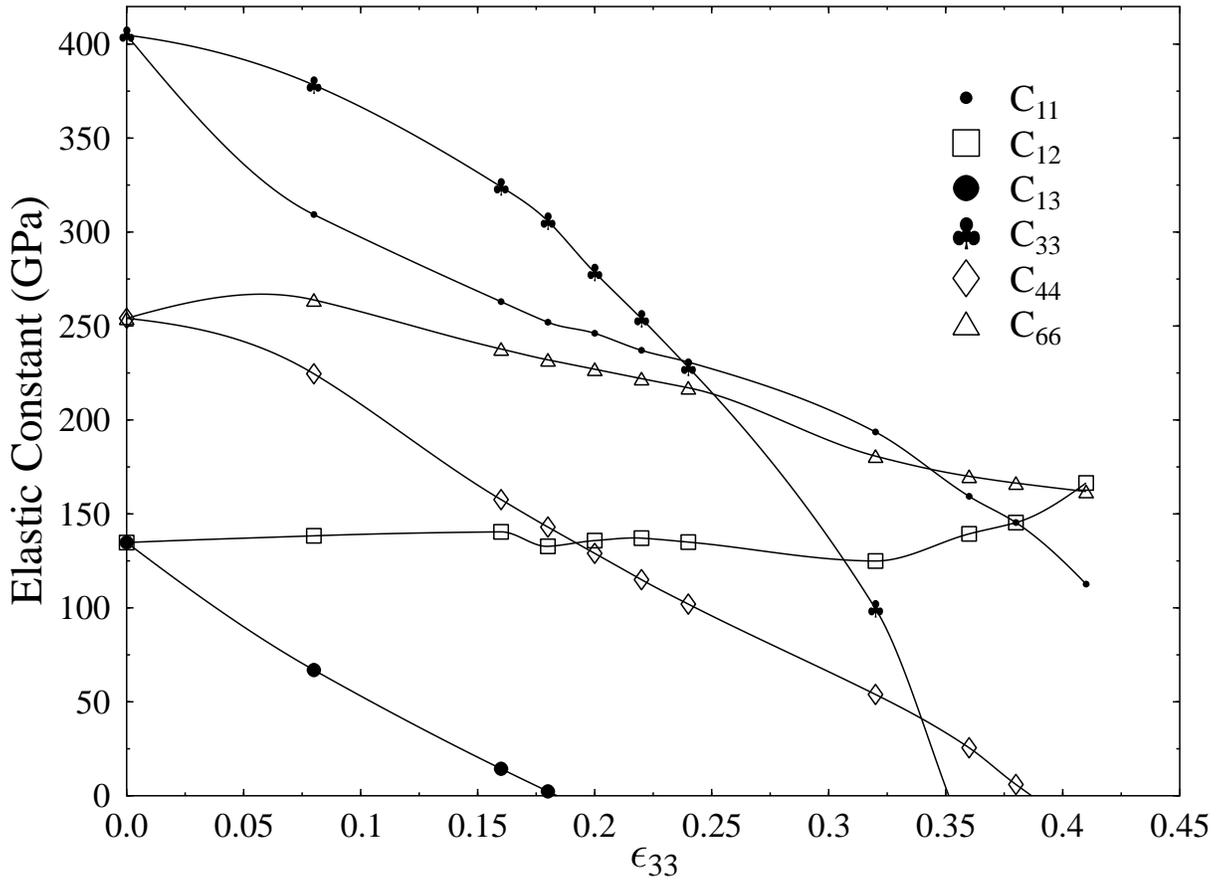
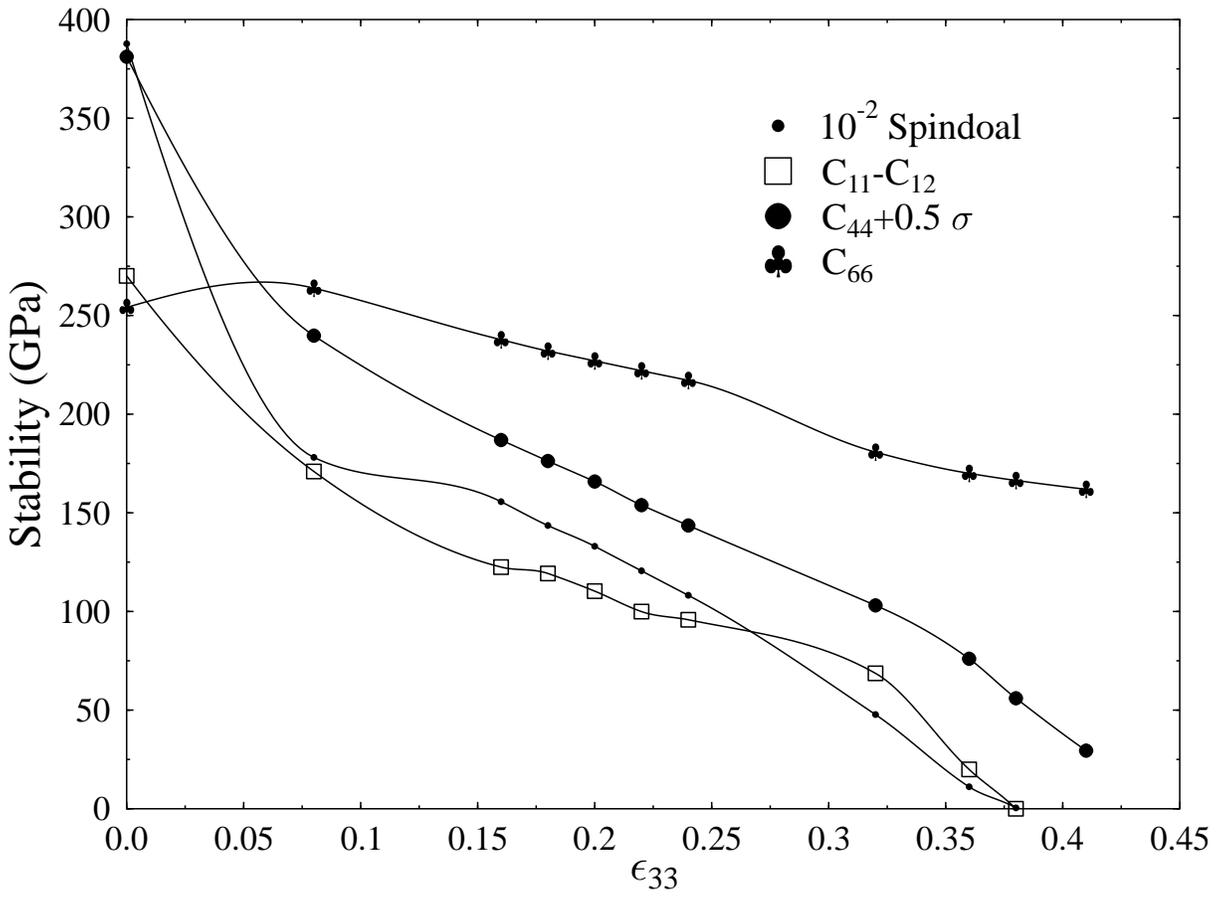
\end{document}